\documentclass{PoS}

\usepackage{latexsym}
\usepackage{amsmath,amssymb,amsfonts}
\usepackage{bm,bbm}
\usepackage{epsfig,graphicx}

\title{Loop functions at finite temperature from perturbation theory to lattice QCD}
\ShortTitle{Loop functions at finite temperature from perturbation theory to lattice QCD}

\author{\speaker{Antonio Vairo}\\
        Physik-Department, Technische Universit\"at M\"unchen, 
        James-Franck-Str. 1, 85748 Garching, Germany\\
        E-mail: \email{antonio.vairo@tum.de}}

\abstract{We review our present knowledge of the Polyakov loop, the correlator of Polyakov loops and the singlet correlator 
          in thermal QCD from the point of view of perturbation theory and lattice QCD.}

\FullConference{XIII Quark Confinement and the Hadron Spectrum - Confinement2018\\
		31 July - 6 August 2018\\
		Maynooth University, Ireland}

\begin{document}

\section{Motivations}
The medium created in heavy-ion collisions and the hard probes chosen to study it are characterized by several energy scales. 
These energy scales and the dynamics of the related degrees of freedom make the study of the full system extremely challenging.

The medium is characterized by several thermodynamical scales that identify different regimes: 
quark-gluon plasma formation, electrical screening, magnetic screening, and others.
Thermal loop functions are quantities suited to assess the size of some of these scales on the lattice.
Indeed, comparing thermal loop functions in perturbation theory with lattice results 
allows to establish at which temperatures (if any, as $g \sim 1$) we may rely on weak-coupling calculations,
and at which temperatures not, at which temperatures and distances electrical screening sets in, 
at which temperatures and distances other screening masses become relevant, and so on.

Quarkonia are possible hard probes of the medium.  A scale that characterizes quarkonia is their size.
Quarkonia of different sizes will react differently to the medium.
Loop functions that are correlators of gauge fields separated by a distance $r$ describe static quark-antiquark pairs.
On the lattice we may vary $r$ and the temperature $T$ in a controlled way establishing what is the relevant interaction at any $r$ and $T$.
Another scale is the binding energy, which for static $Q\bar{Q}$ pairs amounts to the static potential.

For long time the in medium static potential was identified with some free energies defined from suitable correlators. 
This is no more so: the potential describes the real-time evolution of the $Q\bar{Q}$ pair, 
which is, in general, not the case for the free energies; it also has an imaginary part coming from the quarkonium dissociation through scattering 
with the partons in the medium~\cite{Laine:2006ns,Beraudo:2007ky,Brambilla:2008cx}.
Nowadays, the real and imaginary real-time potentials are computed on the lattice by dedicated studies~\cite{Burnier:2014ssa,Burnier:2015tda,Burnier:2016mxc}.
Nevertheless, free energies still offer a clean and controlled setting where to study the thermal dynamics of a static $Q\bar{Q}$ pair.
Moreover, at least the singlet free energy provides a good approximation of the real part of the real-time static potential.

Finally, relevant degrees of freedom describing the real-time evolution of a $Q\bar{Q}$ pair in a medium are 
$Q\bar{Q}$ pairs in a color singlet (bound) and color octet (unbound) configuration, 
reflecting the color decomposition $3 \otimes \bar{3} = 1 \oplus 8$~\cite{Brambilla:2008cx}.
Color octet $Q\bar{Q}$ states are relevant for the out of equilibrium evolution of the $Q\bar{Q}$ pair in the medium through the 
reactions $(Q\bar{Q})_1 \longleftrightarrow  (Q\bar{Q})_8 + \,\textit{gluons}$~\cite{Brambilla:2016wgg,Brambilla:2017zei}.
Thermal loop correlators allow to access also $Q\bar{Q}$ color octet degrees of freedom.

\section{Loop functions}
Loop functions are quantities that can be computed in lattice QCD and that are relevant for the dynamics of static sources in a thermal bath at a temperature $T$~\cite{Petreczky:2005bd}.
Several loop functions have been defined and measured over time. Here, we will focus on the following three.

\begin{figure}[ht]
  \centering
  \includegraphics[width=.24\linewidth]{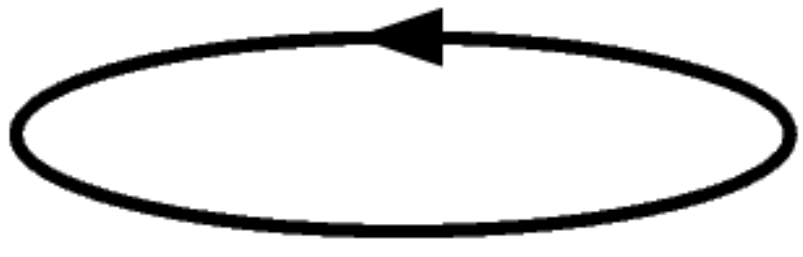}
  \caption{Polyakov loop.}
\label{fig:poly}
\end{figure}

The \textit{Polyakov loop} average in a thermal ensemble at a temperature $T$, depicted in Fig.~\ref{fig:poly}, is defined as  
\begin{equation}
P(T)|_R \equiv \frac{1}{d_R}\langle \textrm{Tr}\, L_R \rangle, \qquad P(T)|_F = e^{-F_Q/T},
\label{eqP}
\end{equation}
where $R$ stands for the color representation of the gauge fields, $d_A=N^2-1$, $d_F=N$, $N=3$ is the number of colors, the trace is over the color matrices, and  
\begin{equation}
L_R({\bf x}) =  \textrm{P} \,\exp \left( ig \int_0^{1/T}d\tau \; A^0({\bf x},\tau) \right), 
\end{equation}
is a straight Wilson line spanning from the Euclidean time $0$ to $1/T$ at the position ${\bf x}$;
P stands for path ordering of the color matrices.
The quantity $F_Q$ is the free energy associated to the Polyakov loop in the fundamental representation, or, in physical terms, to a static quark $Q$.

\begin{figure}[h]
  \centering
  \includegraphics[width=.24\linewidth]{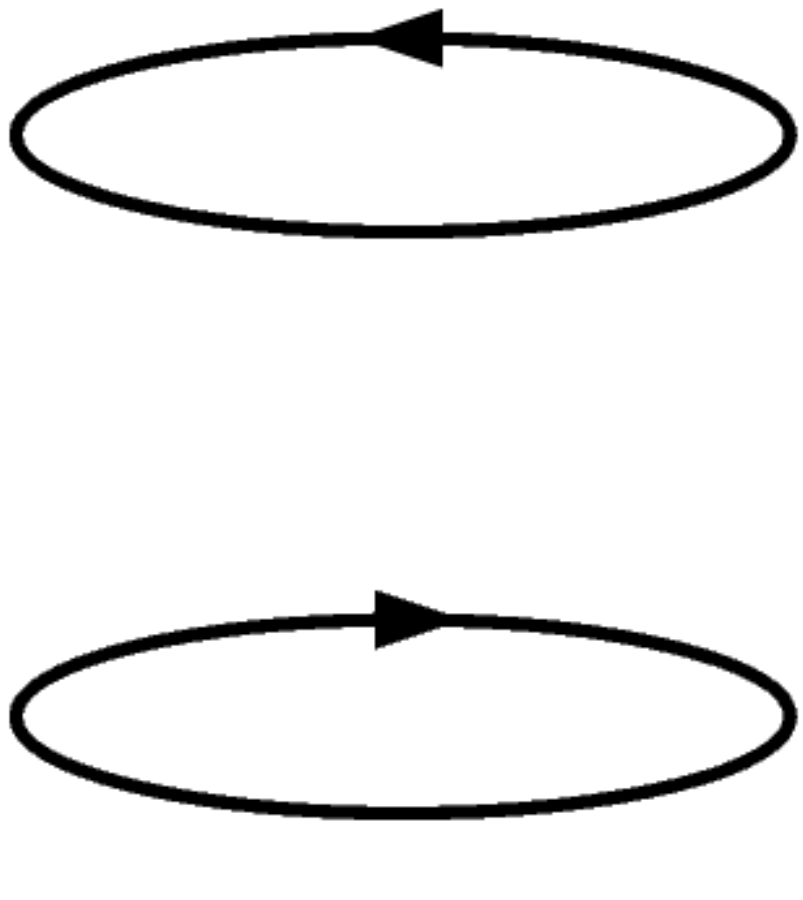}
  \caption{Polyakov loop correlator.}
\label{fig:cor}
\end{figure}

The thermal average of two Polyakov loops separated by a distance ${\bf r}$, shown in Fig.~\ref{fig:cor}, 
is called \textit{Polyakov loop correlator}:
\begin{equation}
P_c(r,T) \equiv \frac{1}{N^2}\langle {\rm Tr} L_F({\bf r})  {\rm Tr} \,L_F^\dagger({\bf 0}) \rangle = e^{-F_{Q\bar{Q}}/T},
\label{eqPc}
\end{equation}
where we have taken the Polyakov loops in the fundamental representation. 
Also in this case we can define a free energy, $F_{Q\bar{Q}}$, which is the free energy of a static quark-antiquark ($Q\bar{Q}$) pair.

\begin{figure}[h]
  \centering
  \includegraphics[width=.24\linewidth]{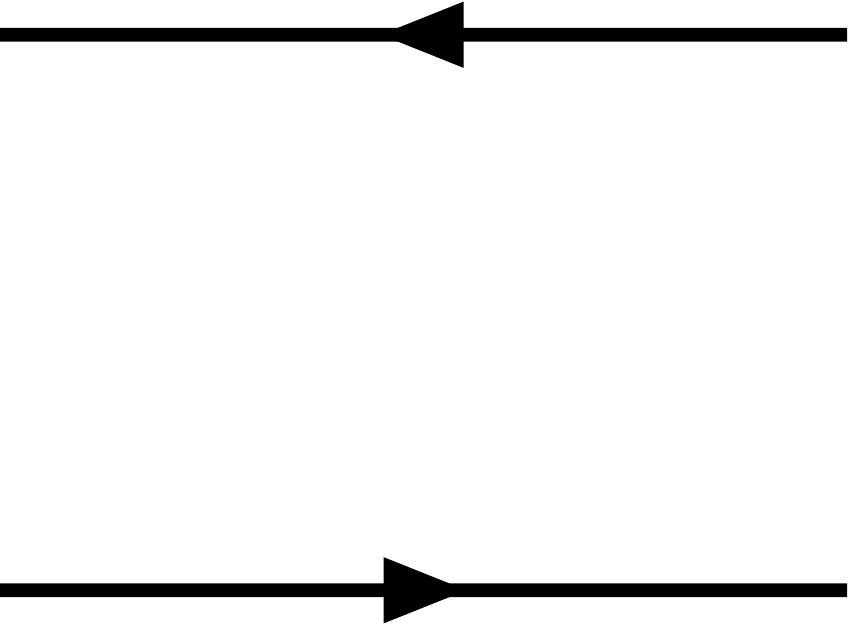}
  \caption{Singlet correlator.}
\label{fig:wi-coul}
\end{figure}

If we consider the thermal average of the trace of two Wilson lines separated by a distance ${\bf r}$ and spanning from $0$ to $1/T$, see Fig.~\ref{fig:wi-coul}, 
then we consider a gauge dependent quantity. 
In Coulomb gauge, for Wilson lines in the fundamental representation, it is called the \textit{singlet correlator}:
\begin{equation}
W_s(r,T) \equiv \frac{1}{N}\langle{\rm Tr} \, L_F({\bf r}) L_F^\dagger({\bf 0}) \rangle = e^{-F_S/T}.
\label{eqWs}
\end{equation}
The corresponding free energy is the singlet free energy, $F_S$.

Other correlators, like the gauge invariant cyclic Wilson loop, have been discussed and computed 
in the literature, see e.g.~\cite{Burnier:2009bk,Berwein:2012mw,Bazavov:2018wmo}, but will not be analyzed further here.
Extensive discussions on the renormalization of the loop functions, which will also not be discussed here,
can be found in~\cite{Brandt:1981kf,Berwein:2013xza,Vairo:2014qma} and references therein.

\section{Polyakov loop}
\label{secpol}
The Polyakov loop enters in all quark-antiquark correlators and free energies.
It also allows to determine the crossover temperature to the quark-gluon plasma, $T_c$.

The Polyakov loop average in the color representation $R$ up to ${\cal O}(g^3)$ is given by: 
\begin{equation}
P|_R = 1 + \frac{C_R\alpha_\mathrm{s}m_D}{2T}, 
\end{equation}
where $C_R$ is the Casimir of the representation $R$ ($C_F=(N^2-1)/(2N)$, $C_A=N$). 
The Debye mass, $m_D$, is an effective mass for temporal modes dynamically generated by the plasma: 
\begin{equation}
\Pi_{00}(k^0=0,|\mathbf{k}|\ll T)  \approx m_D^2 = \frac{2N + n_f}{6}g^2T^2,
\end{equation}
where $\Pi_{00}$ is the temporal component of the gluon self energy and $n_f$ the number of light (massless) flavors.
In a weakly coupled plasma, one typically assumes the hierarchy of energy scales 
$\pi T \gg m_D \sim gT \gg m_M \sim g^2T$, where $m_M$ is the magnetic mass.
Beyond ${\cal O}(g^3)$, it is convenient to compute the Polyakov loop average as the exponential of a reduced number of diagrams
\begin{align}
P|_R&=1\!+\!C_R~\begin{minipage}[b]{0.1\linewidth}\includegraphics[width=\linewidth]{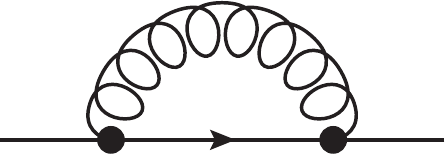}\end{minipage}
+C_R^2~\begin{minipage}[b]{0.1\linewidth}\includegraphics[width=\linewidth]{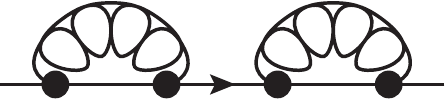}\end{minipage}
+C_R\left(C_R-\frac{C_A}{2}\right)\begin{minipage}{0.1\linewidth}\includegraphics[width=\linewidth]{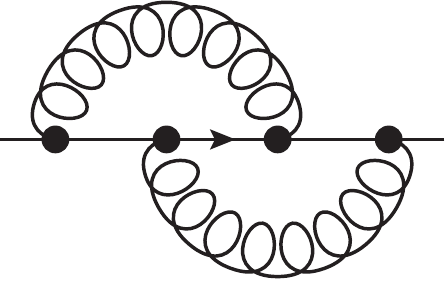}\end{minipage}
+C_R^2~\begin{minipage}[b]{0.1\linewidth}\includegraphics[width=\linewidth]{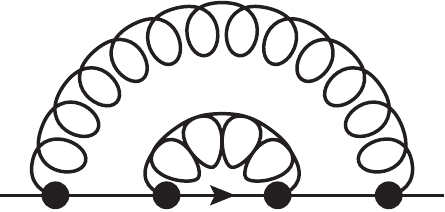}\end{minipage}+\dots\notag\\
 &=\exp\left[C_R~\begin{minipage}[b]{0.1\linewidth}\includegraphics[width=\linewidth]{Poly11.pdf}\end{minipage}
-\frac{1}{2}C_RC_A~\begin{minipage}{0.1\linewidth}\includegraphics[width=\linewidth]{Poly13.pdf}\end{minipage}+\dots\right],
\end{align}
where the dots stand for contributions of order $g^6$ or higher.
We see that at the lowest perturbative orders the exponent is proportional to $C_R$, a property that goes under the name of Casimir scaling.

The Polyakov loop average up to ${\cal O}(g^4)$ reads~\cite{Burnier:2009bk,Brambilla:2010xn}:
\begin{equation}
P|_R = 1+\frac{C_R \alpha_{\rm s} }{2}\frac{m_D}{T}+\frac{C_R \alpha_{\rm s}^2}{2}
\left[C_A\left(\ln\frac{m_D^2}{T^2}+\frac{1}{2}\right)-n_f\ln2\right].
\end{equation}
The logarithm, $\ln{m_D^2}/{T^2}$, signals that an infrared divergence at the scale $T$ has canceled against an ultraviolet divergence 
at the scale $m_D$.

Finally, the Polyakov loop average up to ${\cal O}(g^5)$ reads~\cite{Berwein:2015ayt}:
\begin{align}
\ln P|_R=&\,\frac{C_R\alpha_\mathrm{s}(\mu)\, m_D}{2T}
+\frac{C_R\alpha_\mathrm{s}^2}{2}\left[C_A\left(\frac{1}{2}+\ln\frac{m_D^2}{T^2}\right)-n_f\ln2\right]\notag\\
 &+\frac{3C_R\alpha_\mathrm{s}^2m_D}{16\pi T}\left[3C_A+\frac{2}{3}n_f\left(1-4\ln2\right)+2\beta_0\left(\gamma_E+\ln\frac{\mu}{4\pi T}\right)\right]\notag\\
 &-\frac{C_RC_F n_f\alpha_\mathrm{s}^3T}{4m_D}
  -\frac{C_RC_A^2\alpha_\mathrm{s}^3T}{m_D}\left[\frac{89}{48}+\frac{\pi^2}{12}-\frac{11}{12}\ln2\right],
\end{align}
where $\beta_0=(11N-2n_f)/3$.
Owing to the gauge invariance of the Polyakov loop, the result could be checked in several gauges, like 
Feynman gauge, Coulomb gauge, static gauge ($\partial_0A_0 =0$) and phase-space Coulomb gauge~\cite{Andrasi:2003zf}.

\begin{figure}[ht]
  \centering
  \includegraphics[width=.8\linewidth]{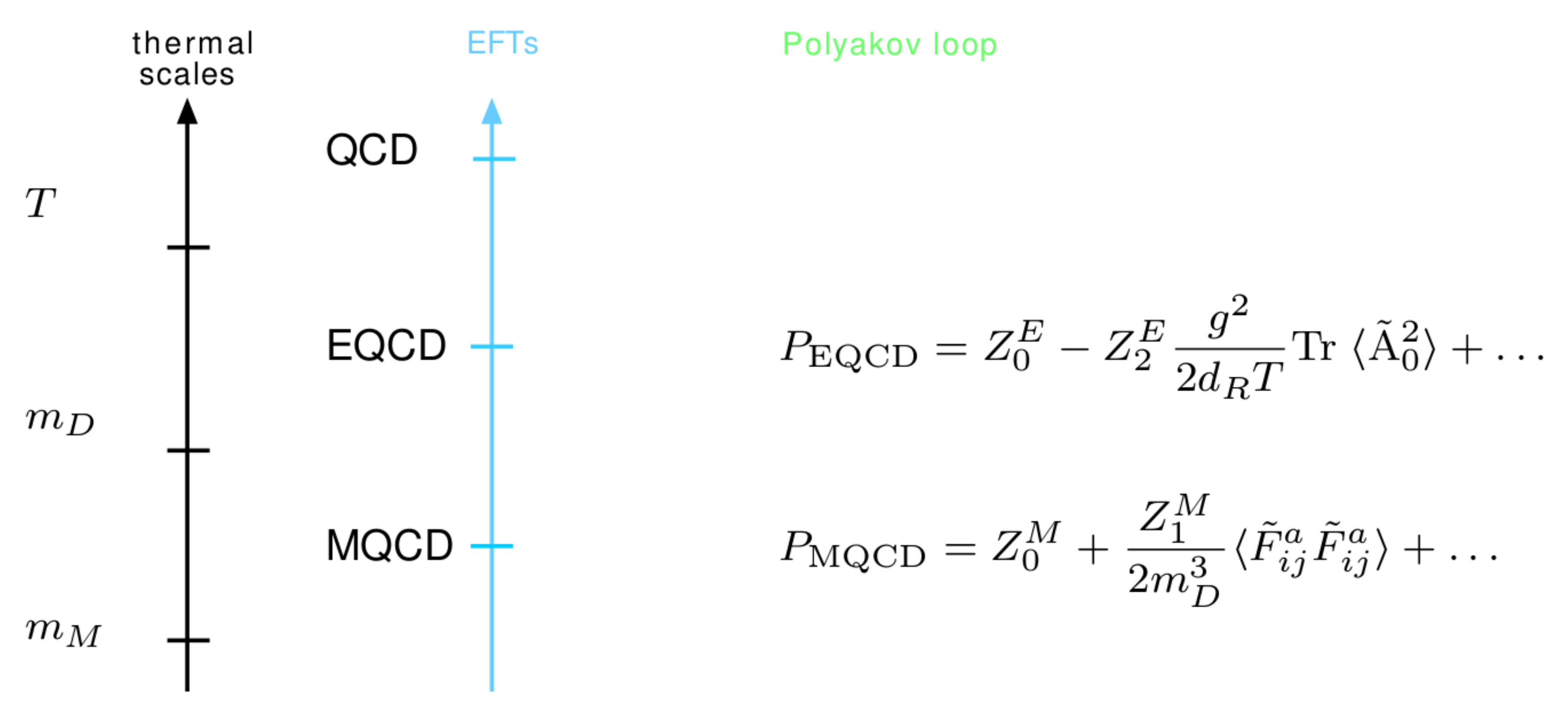}
  \caption{Hierarchy of dimensionally reduced EFTs and Polyakov loop.}
\label{fig:dimred}
\end{figure}

The Polyakov loop average may be computed also using dimensionally reduced effective field theories (EFTs), see Fig.~\ref{fig:dimred}.
In this case, one can take advantage of several existing results, in particular in Refs.~\cite{Braaten:1995jr,Kajantie:1997tt},  
and the calculation up to ${\cal O}(g^5)$ becomes mostly a matter of reorganizing them.
In the framework of dimensionally reduced EFTs it is also straightforward to estimate the order at which 
non-perturbative contributions carried by the magnetic mass, $m_M$, show up first.
In the EFT at the magnetic mass scale (MQCD), the Polyakov loop average gets the contribution $Z_1^M/(2m_D^3) \times \langle \tilde{F}^a_{ij}\tilde{F}^a_{ij}\rangle$, 
which is of order $\alpha_{\rm s}^2/m_D^3 \times m_M^3 \sim g^7$ as $Z_1^M \sim \alpha_{\rm s}^2$. 
This implies that the highest order at which the Polyakov loop average may be computed in perturbation theory is $g^6$, 
which is a computation yet to be done, but in reach of present days techniques.
The order $g^6$ contribution  is the last missing piece of the perturbative expansion of the Polyakov loop.
The assessment of the size of the non-perturbative contributions provides a further test of the full QCD calculation, 
as all Feynman diagrams involving the magnetic mass that could potentially contribute to the Polyakov loop at orders lower than $g^7$ have to cancel.

\begin{figure}[ht]
\centering
 \includegraphics[width=0.7\linewidth]{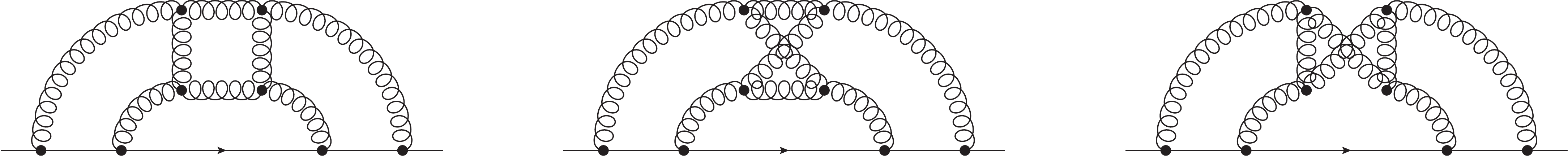}
\caption{Casimir scaling violating diagrams at ${\cal O}(g^8)$.}
\label{fig:casimir}
\end{figure}

By close inspection of the Feynman diagrams contributing to the Polyakov loop, one can show that 
Casimir scaling holds up to ${\cal O}(g^7)$ (including $m_M$ contributions).
Casimir scaling violating diagrams appear first at ${\cal O}(g^8)$.
These are diagrams like the ones shown in Fig.~\ref{fig:casimir}, which contain terms proportional to~\cite{Berwein:2015ayt}  
\begin{equation}
C_R^{(4)}=f^{i_1a_1i_2}\cdots f^{i_4a_4i_1}\frac{1}{d_R}\mathrm{Tr}\left[T_R^{a_1}\cdots T_R^{a_4}\right], 
\qquad {\rm with} \quad 
\frac{C_F^{(4)}}{C_A^{(4)}}=\frac{C_F}{C_A}\frac{N^2+2}{N^2+12}.
\end{equation}
The fact that Casimir scaling is only tinily violated by the Polyakov loop is confirmed also by lattice determinations 
that so far do not show any evidence of violation~\cite{Gupta:2007ax,Petreczky:2015yta}.

\begin{figure}[ht]
\centering
\includegraphics[width=5.8cm]{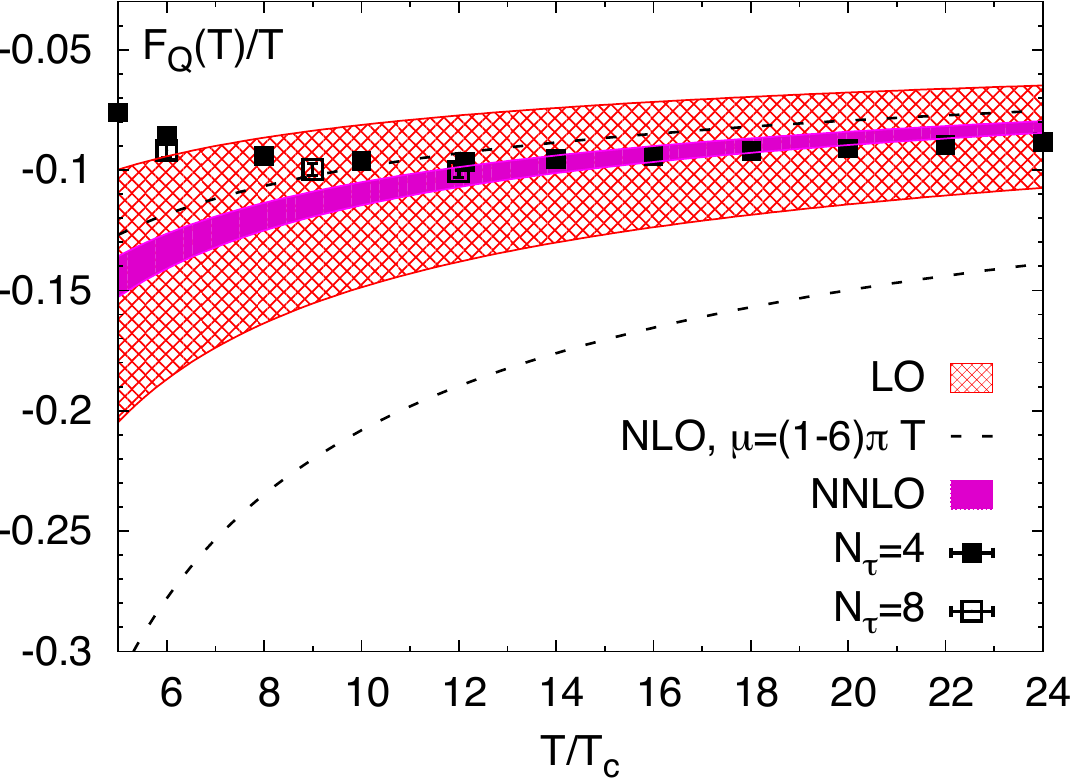}\hspace{2cm}\includegraphics[width=5.8cm]{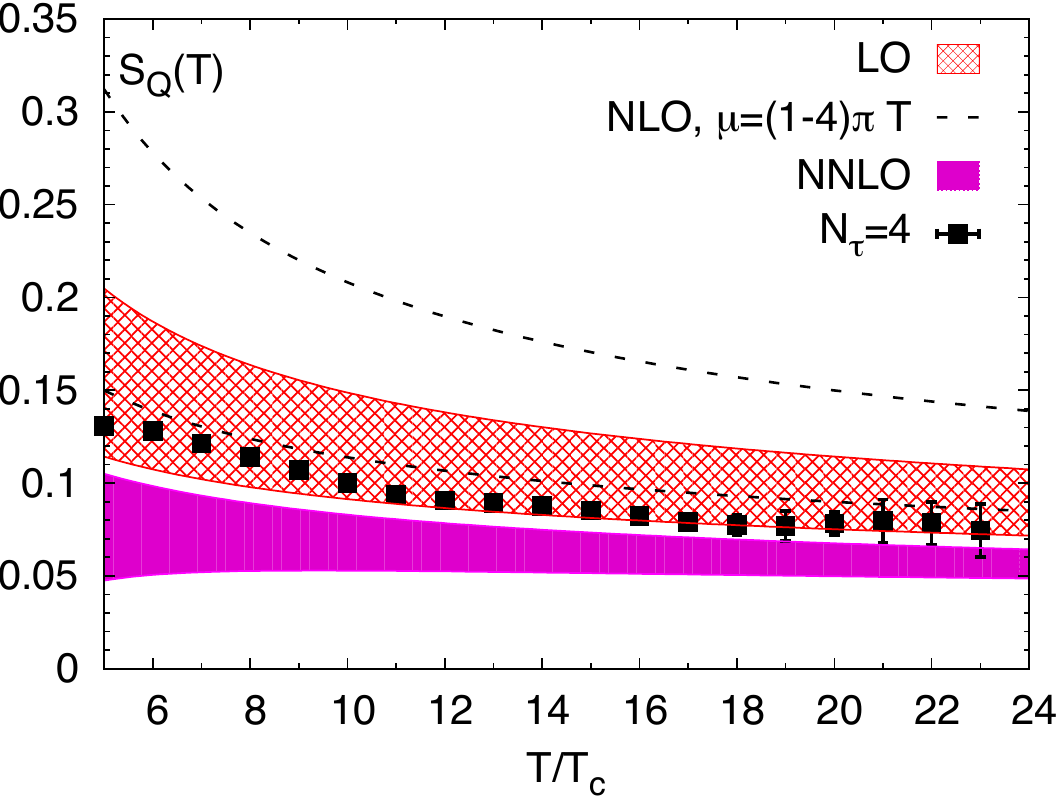}
\caption{Quark free energy (left) and entropy (right) at leading order (LO), next-to-leading order (NLO) and next-to-next-to-leading order (NNLO) 
versus quenched lattice data from~\cite{Gupta:2007ax}. 
The bands are obtained by varying the renormalization scale between $\pi T$ and $6\pi T$.
From~\cite{Berwein:2015ayt}.}
\label{fig:FQ-SQ-quenched}
\end{figure}

In Fig.~\ref{fig:FQ-SQ-quenched} we compare the perturbative expansion of the quark free energy, $F_Q$, and  entropy, $S_Q$, 
\begin{equation}
S_Q = - \frac{\partial F_Q(T)}{\partial T},
\end{equation}
with quenched lattice data. The temperature $T$ is related to the number of temporal lattice steps, $N_\tau$, and the lattice spacing, $a$, through $aN_\tau = 1/T$.
The entropy is a useful quantity to look at, as it does not depend on the normalization shift.
We see an overall good agreement between perturbation theory and lattice data. 
Nevertheless, in particular the entropy data show also a possible sensitivity to the missing order $g^6$ contribution at high temperatures.

\begin{figure}[ht]
\makebox[0cm]{\phantom b}
\centering
\includegraphics[width=6.5cm]{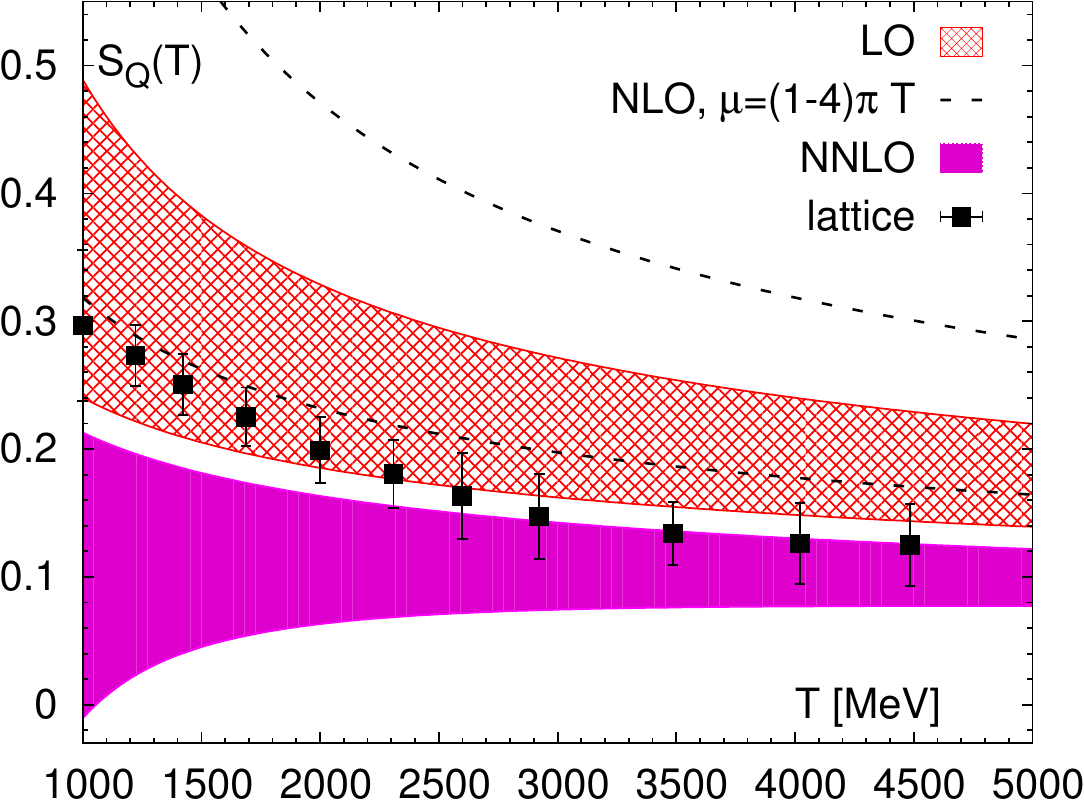}
\caption{Entropy at LO, NLO and NNLO versus 2+1 flavor QCD lattice data with $N_\tau=4$.
The bands are obtained by varying the renormalization scale between $\pi T$ and $4\pi T$.
From~\cite{Bazavov:2016uvm}.}
\label{fig:SQ}
\end{figure}

In Fig.~\ref{fig:SQ} we compare the perturbative expansion of the quark entropy, $S_Q$, with 2+1 flavor QCD lattice data.
Again, we observe good agreement between perturbation theory and lattice data at high temperatures.
At low temperatures (not in the plot) the entropy shows a peak. 
The position of the entropy peak is at $T_S = 153^{+6.5}_{-5}$ MeV, a value that is consistent with the crossover temperature to the quark-gluon plasma~\cite{Borsanyi:2010bp,Bazavov:2011nk}.

\section{Polyakov loop correlator}
The correlator of two Polyakov loops located at a distance ${\bf r}$ allows to compute the free energy of a static $Q\bar{Q}$ pair, $F_{Q\bar{Q}}$. 
Eventually, effective field theories allow to link the correlator also to suitably defined $Q\bar{Q}$ color singlet and color octet free energies. 

The correlator of two Polyakov loops and the corresponding $Q\bar{Q}$ free energy, see Eq. \eqref{eqPc}, depend on several energy scales: 
$1/r$, $\alpha_{\rm s}/r$, ..., $\pi T$, $m_D$, ... . 
A strict perturbative expansion, i.e. a series in powers of $g$, requires $\pi T \gg m_D \gg \alpha_{\rm s}/r$. 
Moreover we may require, $1/r \gg \pi T$, which implies that the non-perturbative magnetic mass is parametrically smaller than all above scales and, 
in particular, smaller than the Coulomb potential, $\alpha_{\rm s}/r$. The condition $1/r \gg \pi T$ restricts the validity of the results to short distances.
Under these conditions the Polyakov loop correlator, once divided by the Polyakov loop average squared, reads up to order $g^6$~\cite{Berwein:2017thy}
\begin{align}
  \exp\left[\frac{2F_Q-F_{Q\bar{Q}}}{T}\right]_{{\rm up~to~}g^6}
  ={}&1+\frac{N^2-1}{8N^2}\left\{\frac{\alpha_\mathrm{s}^2(1/r)}{r^2T^2} - \frac{2\alpha_\mathrm{s}(1/r)\alpha_\mathrm{s}(4\pi T)m_D(4\pi T)}{rT^2}\right.\notag\\
  &\hspace{-1cm}
  + \frac{N^2-2}{6N}\frac{\alpha_\mathrm{s}^3(1/r)}{r^3T^3}  
  + \frac{\alpha_\mathrm{s}(1/r)\alpha_\mathrm{s}^2}{2\pi r^2T^2}\left(\frac{31}{9}N-\frac{10}{9}n_f+2\beta_0\gamma_E\right)\notag\\
  &\hspace{-1cm}
  +\frac{2\alpha_\mathrm{s}(1/r)\alpha_\mathrm{s}^2}{rT}\left[N\left(1-\frac{\pi^2}{8}+\ln\frac{T^2}{m_D^2}\right)+n_f\ln2\right]\notag\\
  &\hspace{-1cm}
  -\frac{2\pi N\alpha_\mathrm{s}(1/r)\alpha_\mathrm{s}^2}{9}+\frac{\alpha_\mathrm{s}^2(4\pi T)m_D^2(4\pi T)}{T^2}
  +2\alpha_\mathrm{s}(1/r)\alpha_\mathrm{s}^2\left(\frac{4}{3}N+n_f\right)\zeta(3)rT\notag\\
  &\hspace{-1cm}
  -\left.2\pi\alpha_\mathrm{s}(1/r)\alpha_\mathrm{s}^2\left(\frac{22}{675}N+\frac{7}{270}n_f\right)(r\pi T)^2\right\}+\mathcal{O}\left(g^6(r\pi T)^4\right),
\label{Pcg6}
\end{align}
and at order $g^7$ 
\begin{align}
 \exp\left[\frac{2F_Q-F_{Q\bar{Q}}}{T}\right]_{g^7}={}&\frac{N^2-1}{8N^2}\left\{-\frac{N^2-2}{2N}\frac{\alpha_\mathrm{s}^2(1/r)\alpha_\mathrm{s}(4\pi T)m_D(4\pi T)}{r^2T^3}\right.\notag\\
 &\hspace{-1.5cm}
 -\frac{2\alpha_\mathrm{s}^2\alpha_\mathrm{s}(4\pi T)m_D(4\pi T)}{4\pi rT^2}\left(\frac{31}{9}N-\frac{10}{9}n_f+2\beta_0\gamma_E\right)\notag\\
 &\hspace{-1.5cm}
 -\frac{3\alpha_\mathrm{s}(1/r)\alpha_\mathrm{s}^2m_D}{4\pi rT^2}\left[3N+\frac{2}{3}n_f(1-4\ln2)+2\beta_0\gamma_E\right]\notag\\
 &\hspace{-1cm}
 +\frac{(N^2-1)n_f}{2N}\frac{\alpha_\mathrm{s}(1/r)\alpha_\mathrm{s}^3}{rm_D}+\frac{2N^2\alpha_\mathrm{s}(1/r)\alpha_\mathrm{s}^3}{rm_D}\left[\frac{89}{24}+\frac{\pi^2}{6}-\frac{11}{6}\ln2\right]\notag\\
 &\hspace{-1.5cm}
 -\frac{2\alpha_\mathrm{s}^2\alpha_\mathrm{s}(4\pi T)m_D(4\pi T)}{T}\left[N\left(-\frac{1}{2}+\ln\frac{T^2}{m_D^2}\right)+n_f\ln2\right]\notag\\
 &\hspace{-1.5cm}
 -\frac{\alpha_\mathrm{s}(1/r)\alpha_\mathrm{s}m_D^3}{3T^3}rT+\frac{2\pi N\alpha_\mathrm{s}^2\alpha_\mathrm{s}(4\pi T)m_D(4\pi T)}{9T}rT\notag\\
 &\hspace{-1.5cm}
 -\frac{2\alpha_\mathrm{s}^2\alpha_\mathrm{s}(4\pi T)m_D(4\pi T)}{T}\left(\frac{4}{3}N+n_f\right)\zeta(3)(rT)^2\notag\\
 &\hspace{-1.5cm}
 +\left.\frac{2\alpha_\mathrm{s}^2\alpha_\mathrm{s}(4\pi T)m_D(4\pi T)}{T}\left(\frac{22}{675}N+\frac{7}{270}n_f\right)(r\pi T)^3\right\}
 +\mathcal{O}\left(g^7(r\pi T)^4\right).
\label{Pcg7}
\end{align}
The scale of $\alpha_\mathrm{s}$ is an arbitrary renormalization scale $\mu$, if not differently specified.
We notice that in this setting $F_{Q\bar{Q}}$ at leading order is not the Coulomb potential.

\begin{figure}
\centering
\includegraphics[width=7.2cm]{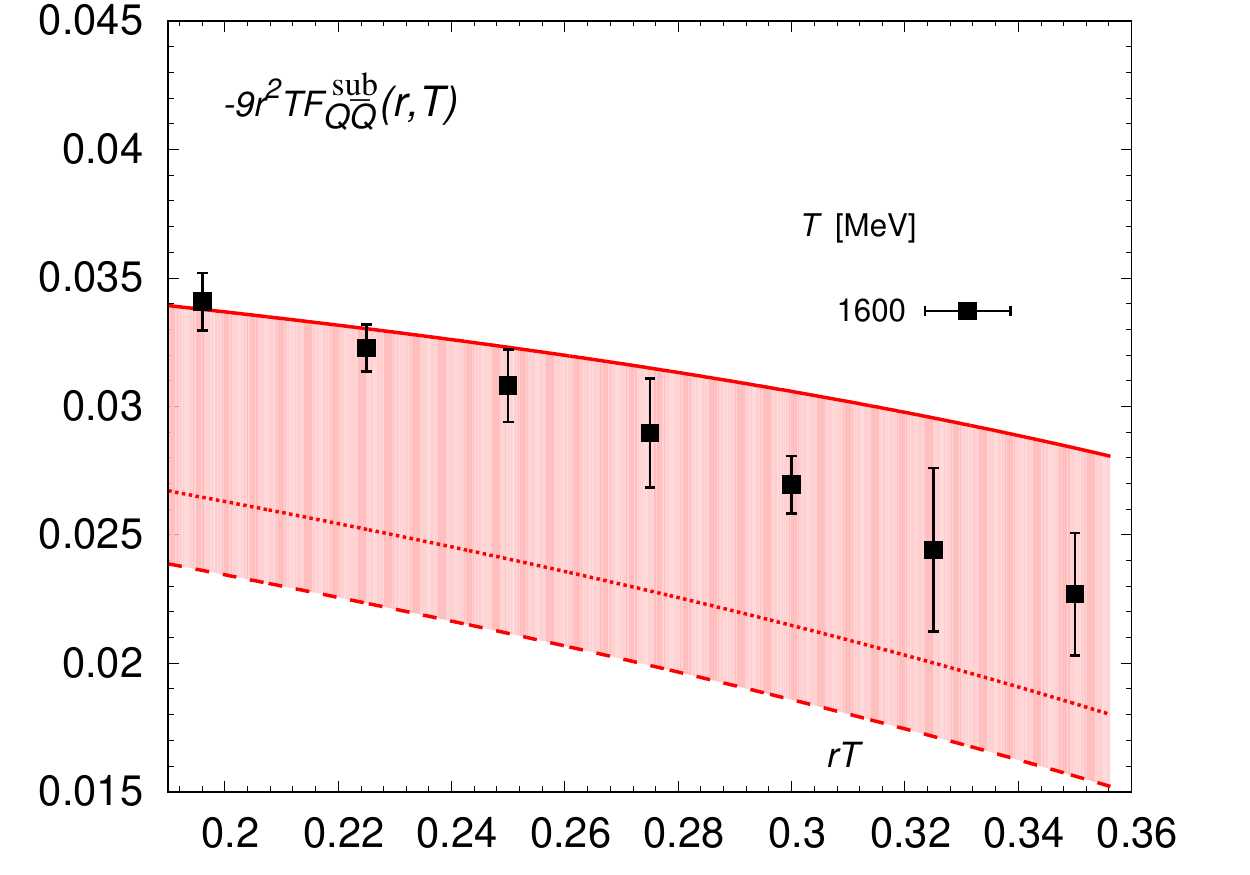}
\caption{The (twice the quark free energy) subtracted $Q\bar{Q}$ free energy at short distances at $T=1600$~MeV, computed on a 2+1 flavor lattice, 
compared to the most accurate perturbative expression given in the text. 
The band follows from varying the renormalization scale from $\pi T$ to $4 \pi T$. From ~\cite{Bazavov:2018wmo}.}
\label{fig:FQQ-short}
\end{figure}

In Fig.~\ref{fig:FQQ-short} we compare $F_{Q\bar{Q}}$ given by the above perturbative expression with 2+1 flavor lattice data at short distances and at a high temperature~\cite{Bazavov:2018wmo}.
We find good agreement.

\begin{figure}
\centering
\includegraphics[width=7.2cm]{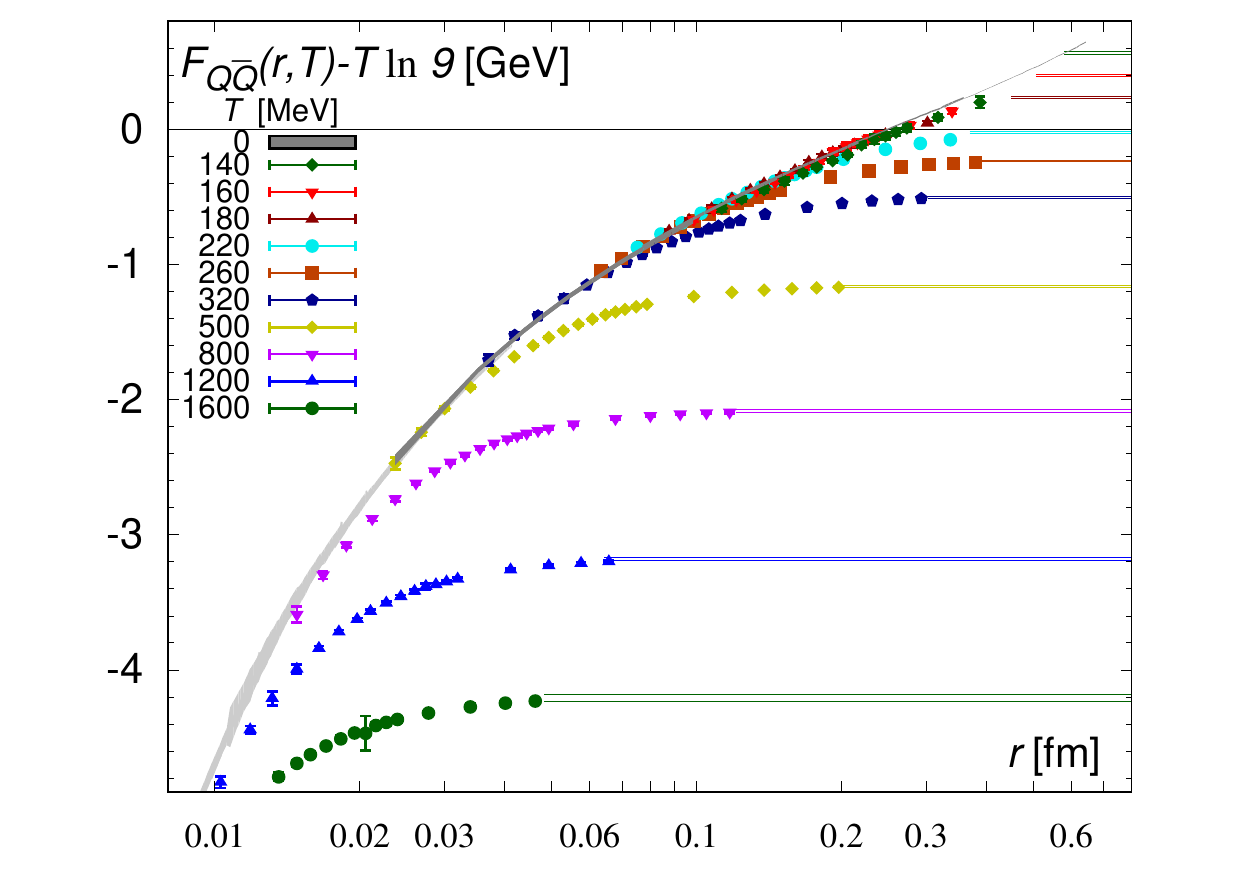}
\caption{The $Q\bar{Q}$ free energy computed on a 2+1 flavor lattice. 
The dark gray band shows the $T=0$ static energy, and the light gray band shows the singlet free energy (see next section) at high temperatures and very short distances.
The horizontal bands correspond to $2 F_Q$. From ~\cite{Bazavov:2018wmo}.}
\label{fig:FQQ}
\end{figure}

However, $F_{Q\bar{Q}}$ has a more complicated behaviour with temperature and distance than the above comparison may suggest.
In Fig.~\ref{fig:FQQ} we show $F_{Q\bar{Q}}$, measured on a 2+1 flavor lattice~\cite{Bazavov:2018wmo}, over a wide range of temperatures and distances.
We see that at short distances and low temperatures $F_{Q\bar{Q}}$ does overlap with the $T=0$ Coulomb potential, 
while the agreement with strict perturbation theory, highlighted in the previous Fig.~\ref{fig:FQQ-short}, requires somewhat higher temperatures.
Eventually, at large distances the correlator goes over to two infinitely separated Polyakov loops.

In order to understand the two regimes of the Polyakov loop correlator at short distances, 
it is useful to look at the correlator in the EFT language of potential non-relativistic QCD (pNRQCD).
In pNRQCD, the Polyakov loop correlator, $P_c(r,T)$, can be put in the form~\cite{Brambilla:2010xn,Berwein:2017thy}
\begin{equation}
P_c(r,T) = \frac{1}{N^2}\Bigg[ e^{-{f_{s}(r,T,m_D)}/T} + (N^2-1)e^{-{f_{o}(r,T,m_D)}/T} + {\cal O}\left(\alpha_{\rm s}^3(r\pi T)^4\right)\Bigg], 
\label{PcpNRQCD}
\end{equation}
where $f_{s}$ is a gauge-invariant $Q\bar{Q}$ color singlet free energy and $f_{o}$ a gauge-invariant $Q\bar{Q}$ color octet free energy. 
The free energies $f_s$ and $f_o$ can be computed from the singlet and octet propagators in pNRQCD:
\begin{align}
\frac{\langle S(\bm{r},{\bf 0},1/T)S^\dagger(\bm{r},{\bf 0},0)\rangle}{\cal N} &=  e^{-V_s(r)/T}(1+ \delta_s) \equiv e^{- f_{s}(r,T,m_D)/T}, \\
\frac{\langle O^a(\bm{r},{\bf 0},1/T)O^{a\,\dagger}(\bm{r},{\bf 0},0)\rangle}{\cal  N}  &=  
e^{-V_o(r)/T}\left[(N^2-1)\, P|_A   +  \delta_o\right]
\equiv (N^2-1)e^{-f_{o}(r,T,m_D)/T} , 
\end{align}
where ${\cal N}$ is a normalization factor, 
$S$ and $O^a$ are $Q\bar{Q}$ fields in a color singlet and octet configuration respectively, 
$V_s = -C_F\alpha_{\rm s}/r + \dots$ and $V_o = \alpha_{\rm s}/(2Nr) + \dots$ are the color singlet and octet, Coulomb-like, static potentials 
respectively, $P|_A$ is the Polyakov loop in the adjoint representation, 
and  $\delta_s$ and $\delta_o$ stand for thermal loop corrections to the singlet and octet propagators.

From Eq.~\eqref{PcpNRQCD} it is clear that we may identify at least two possible regimes:
a low temperature regime, $T \ll V_s$ (or $rT \ll \alpha_{\rm s}$), for which the Polyakov loop correlator may  be approximated 
by the color singlet Coulomb potential, $P_c \approx {e^{-{V_s/T}}}/{N^2}$. This is the gray band of Fig.~\ref{fig:FQQ}.
But we may also identify a high temperature regime, $T \gg V_s$ (or $rT \gg \alpha_{\rm s}$), 
for which $P_c$ is a linear combination of  $e^{-{f_s/T}}$ and $e^{-{f_o/T}}$.
The strict perturbative expansion in $g$ that led to Eqs.~\eqref{Pcg6} and \eqref{Pcg7} corresponds to this last regime.

\begin{figure}
\centering
\includegraphics[width=7.2cm]{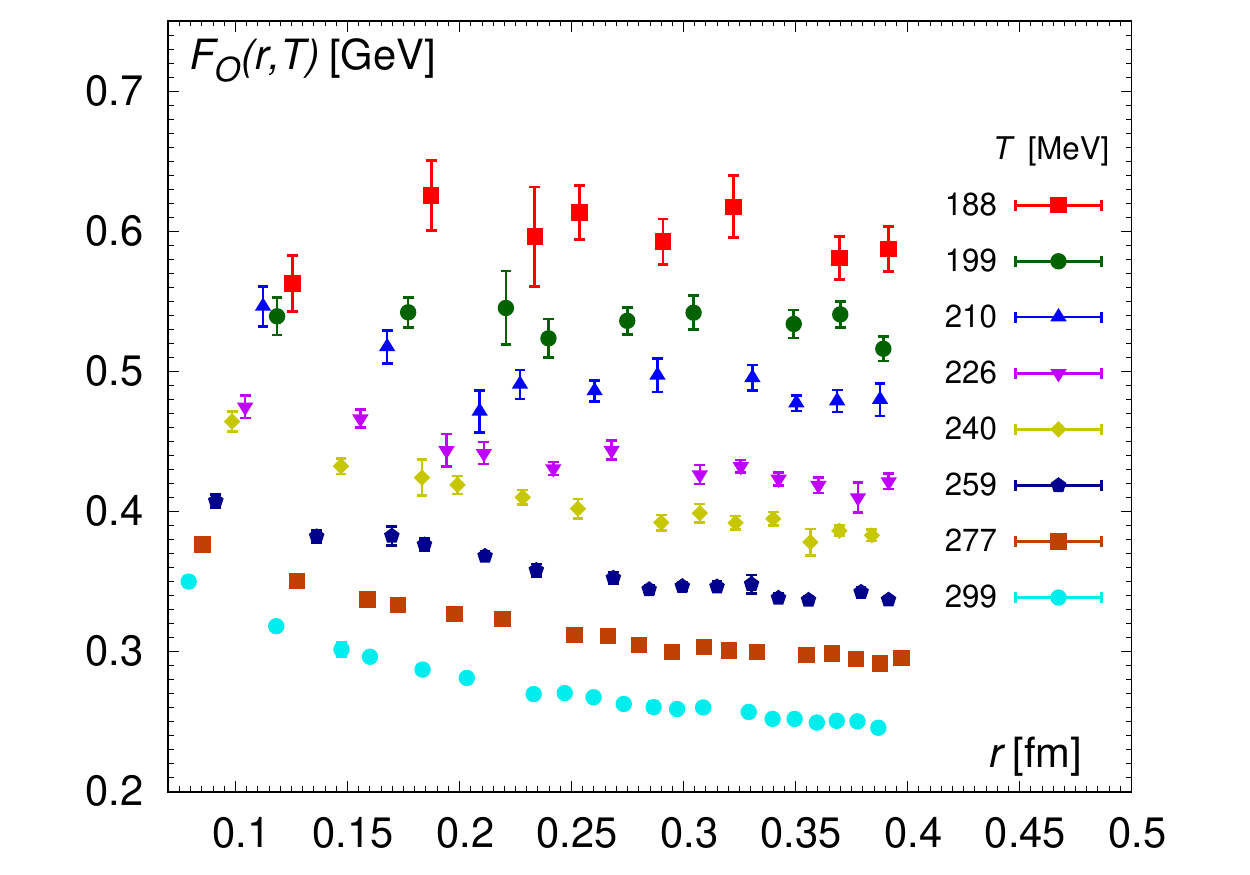}
\caption{The octet free energy determined on a 2+1 flavor lattice with $N_{\tau}=8$ at different distances and temperatures. From~\cite{Bazavov:2018wmo}.}
\label{fig:octet-lat}
\end{figure}

\begin{figure}
\centering
\includegraphics[width=6cm]{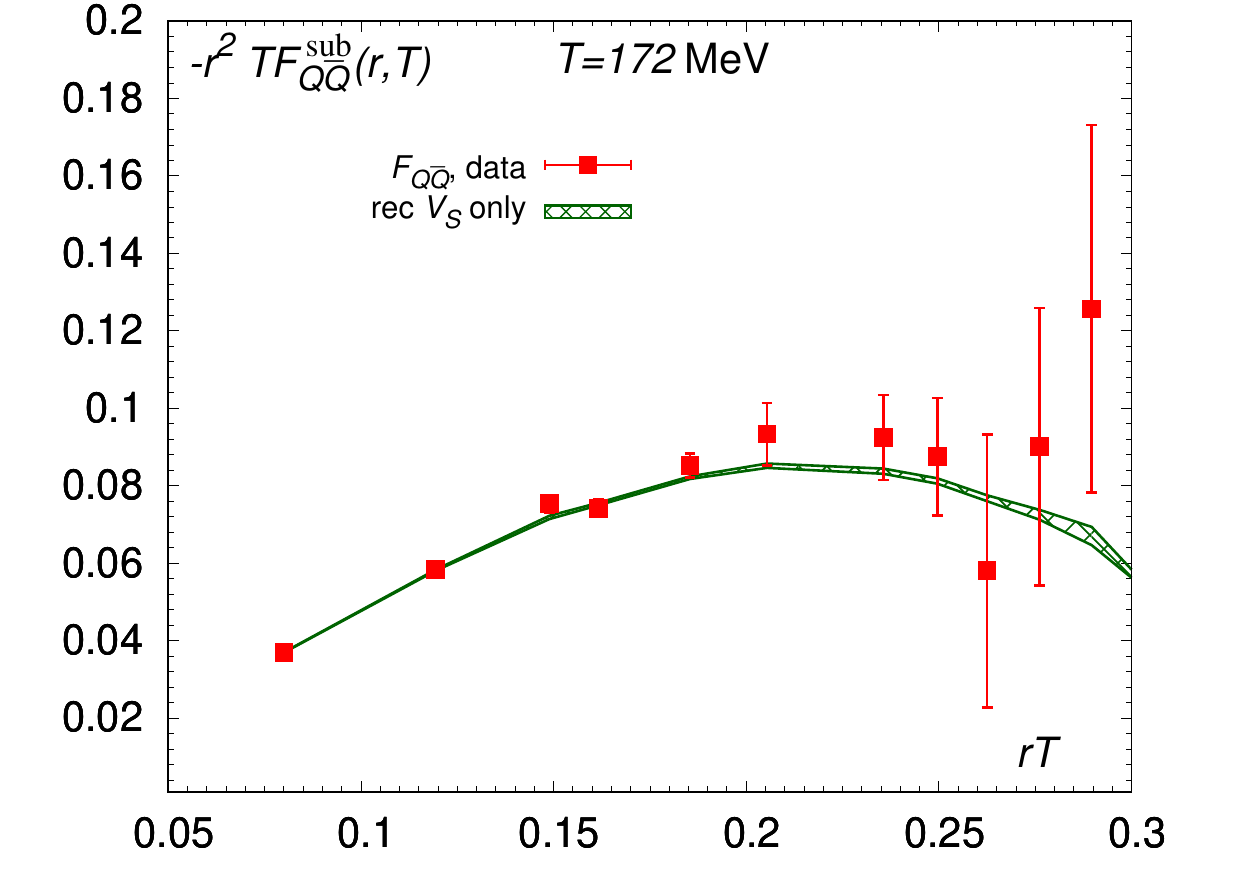}\hspace{2cm}\includegraphics[width=5.8cm]{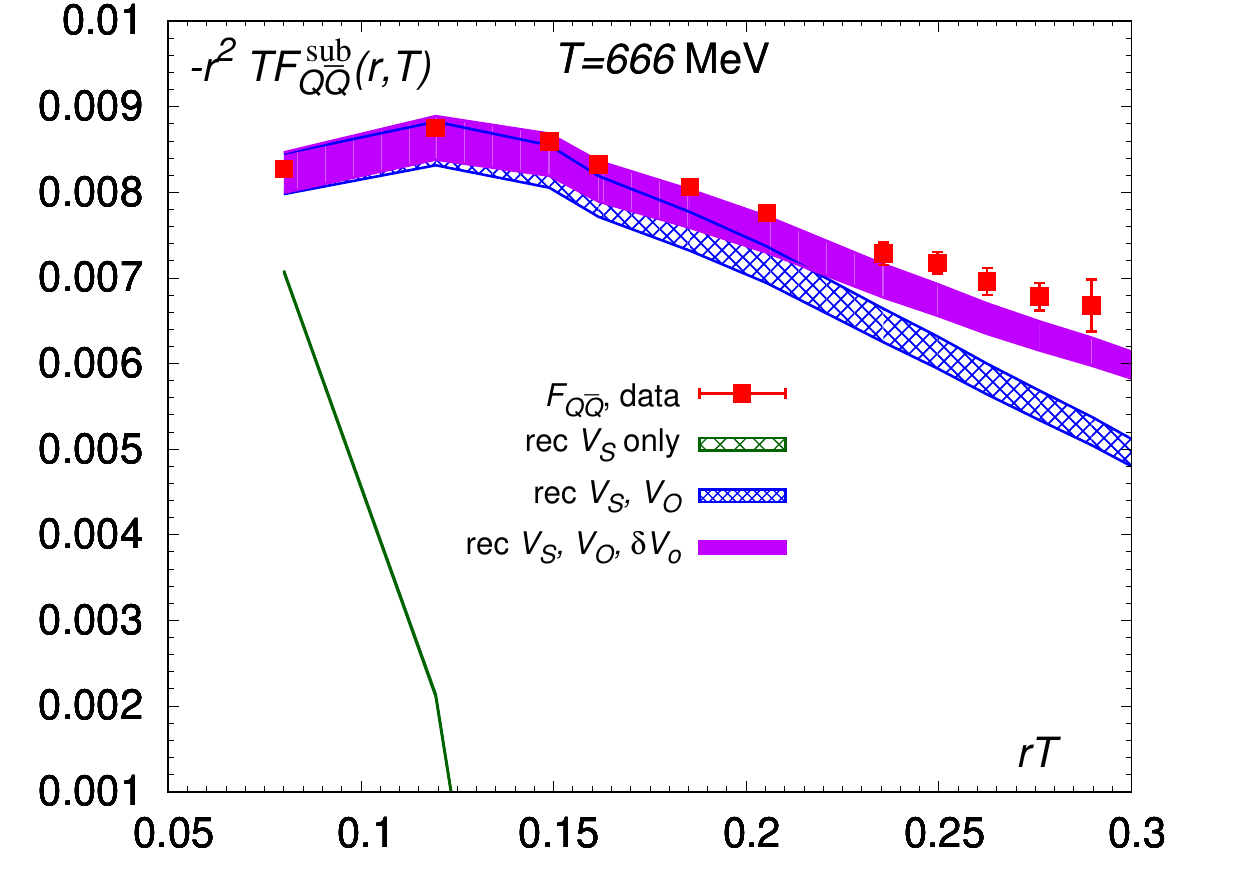}
\caption{The (twice the quark free energy) subtracted $Q\bar{Q}$ free energy calculated on a 2+1 flavor lattice with $N_{\tau}=12$ 
compared to the right-hand-side of Eq.~\eqref{PcpNRQCD} at a low temperature (left panel) and at a higher temperature (right panel). 
The quark free energy is taken from lattice results (see Sect.~\ref{secpol}).
The color singlet static potential, $V_s$, as well as the Casimir-scaling part of the color octet static potential, $V_o$, are reconstructed from the static energy computed on the lattice. 
The Casimir scaling violating contribution to the octet potential, $\delta V_o$, comes from perturbation theory. From~\cite{Bazavov:2018wmo}.}
\label{fig:octet-lat-pNRQCD}
\end{figure}

From Eq.~\eqref{PcpNRQCD}, using the $T=0$ lattice static potential as an input for $f_s$ and lattice data for $P_c$, 
one can determine the octet free energy on the lattice, $F_O$, see Fig.~\ref{fig:octet-lat}.
In turn, $F_O$ should provide a good approximation for the pNRQCD color octet free energy, $f_o$.
Indeed, in the short range we see the onset of a repulsive Coulomb potential.
The existence of a low and a high temperature regime in the behaviour of the Polyakov loop correlator, $P_c$,
and the relevance of the octet degrees of freedom is well demonstrated by the two diagrams of Fig.~\ref{fig:octet-lat-pNRQCD}.
In the left panel, we see that at a sufficiently low temperature the data are well described by the color singlet 
static potential alone (taken from $T=0$ lattice data). In the right panel, at a somewhat higher temperature, 
we see that the data not only require a color octet potential, obtained by color rescaling the color singlet static potential, 
but are also sensitive to the two-loop Casimir scaling breaking effects computed in Ref.~\cite{Kniehl:2004rk}.

At very large distances the Polyakov loop correlator gets screened, first by the Debye mass, then by dynamically generated asymptotic screening masses. 
We refer to Ref.~\cite{Bazavov:2018wmo} (and, at this conference, to Ref.~\cite{Steinbeisser:2018sde}) for a complete and updated discussion.

\section{Singlet correlator in Coulomb gauge}
The singlet correlator, Eq.~\eqref{eqWs}, in Coulomb gauge allows to define a singlet free energy, $F_S$.
Its expression up to order $g^4$ at short distances reads~\cite{Burnier:2009bk,Berwein:2017thy}
\begin{align}
 \frac{F_S}{T}={}&-\frac{N^2-1}{2N}\frac{\alpha_\mathrm{s}(1/r)}{rT}\left[1+\frac{\alpha_\mathrm{s}}{4\pi}\left(\frac{31}{9}N-\frac{10}{9}n_f+2\beta_0\gamma_E\right)\right]+\frac{1}{18}\left(N^2-1\right)\alpha_\mathrm{s}^2r\pi T\notag\\
 &-\frac{N^2-1}{2N}\left(\frac{4}{3}N+n_f\right)\zeta(3)\alpha_\mathrm{s}^2r^2T^2+\frac{N^2-1}{12N}\frac{\alpha_\mathrm{s}m_D^3}{T^3}r^2T^2\notag\\
&+\frac{N^2-1}{2 N}\left(\frac{22}{675}N+\frac{7}{270}n_f\right)\alpha_\mathrm{s}^2(r\pi T)^3
+\mathcal{O}\left(\alpha_\mathrm{s}^2(r\pi T)^5,\alpha_\mathrm{s}^3\right) \,.
\label{eq:Fsg4}
\end{align}
\begin{figure}[ht]
\centering
\includegraphics[width=7.2cm]{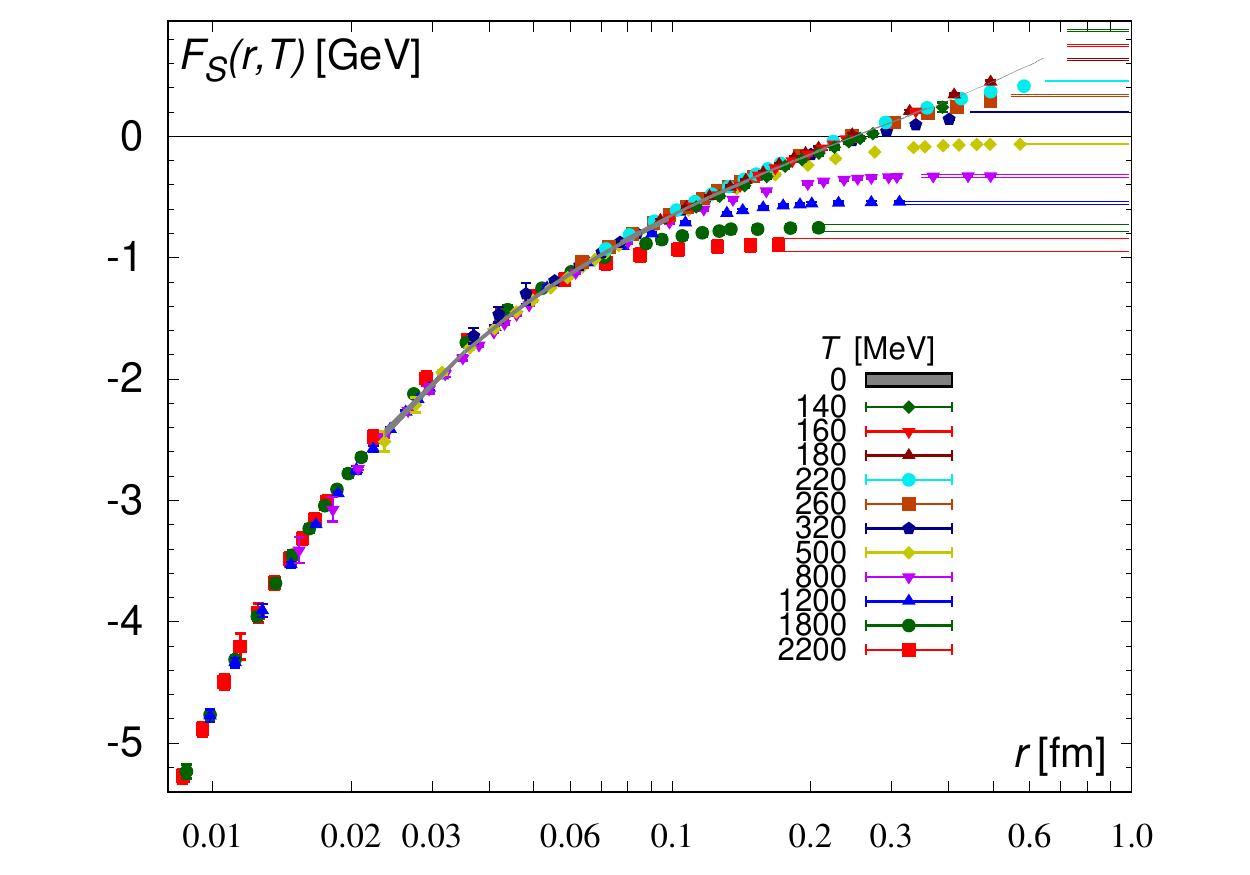}
\caption{The singlet free energy computed on a 2+1 flavor lattice. 
The dark gray band shows the $T=0$ static energy.
The horizontal bands correspond to $2 F_Q$. From ~\cite{Bazavov:2018wmo}.
}
\label{fig:Fs-lattice}
\end{figure}
Similarly one may compute a free energy in the adjoint representation.
The first two lines of Eq.~\eqref{eq:Fsg4} coincide with the real part of the real-time static potential computed in Ref.~\cite{Brambilla:2008cx}.
If this property will also hold at higher orders or if it is an accident valid only at the present accuracy is an open question to be further investigated.

\begin{figure}[ht]
\centering
\includegraphics[width=7.2cm]{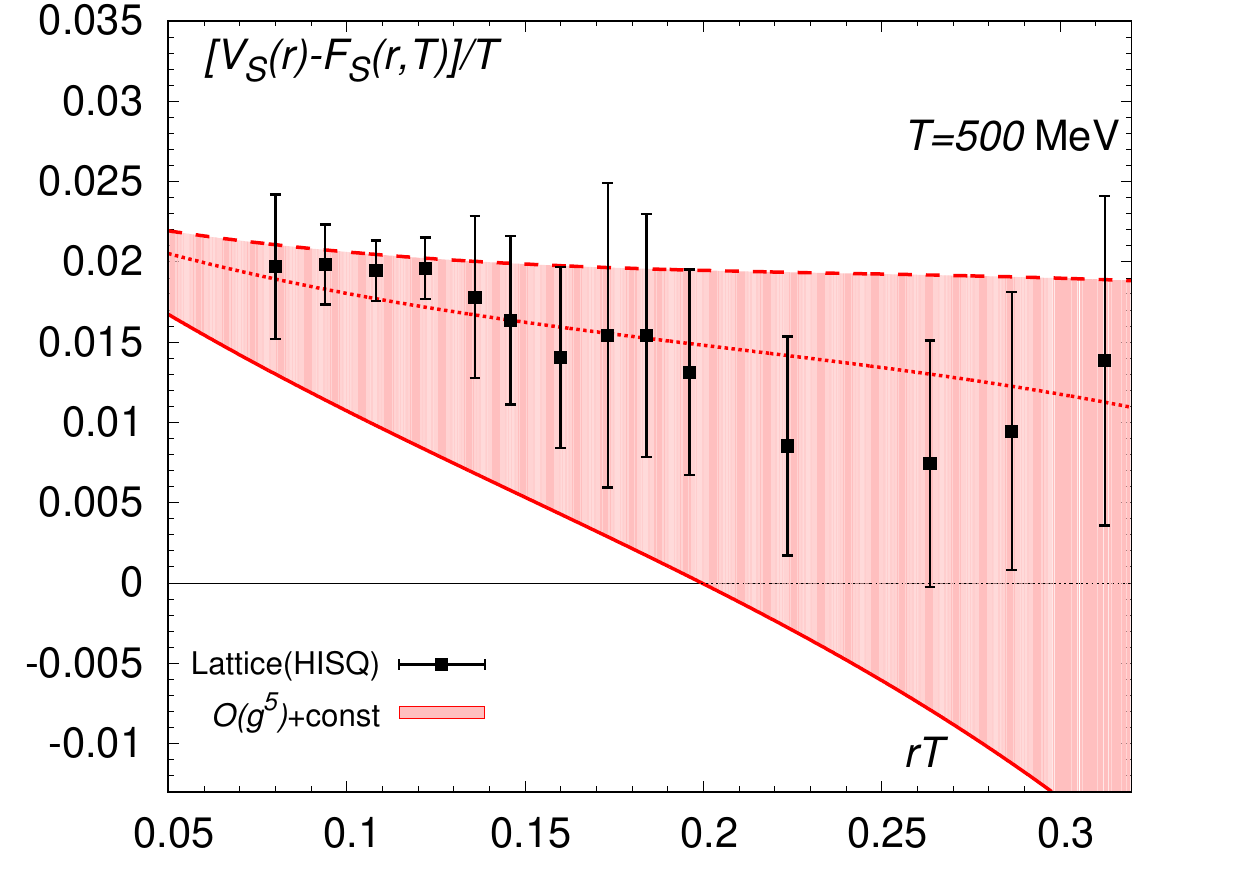}
\caption{The singlet free energy at short distances at $T=500$~MeV subtracted of the $T=0$ static potential, 
computed on a 2+1 flavor lattice, compared to the most accurate perturbative expression given in the text. 
The band follows from varying the renormalization scale from $\pi T$ to $4 \pi T$.
From ~\cite{Bazavov:2018wmo}.}
\label{fig:Fs-short}
\end{figure}

Lattice data for $F_S$ are shown in Fig.~\ref{fig:Fs-lattice}. We see that, because there is no mixing and cancellation with 
the octet potential at leading order, the singlet free energy follows the behaviour of the $T=0$ static potential 
for a wide range of distances and temperatures, before eventually becoming sensitive to the screening induced by the Debye mass.
This behaviour clearly distinguishes the singlet free energy from the $Q\bar{Q}$ free energy discussed in the previous section.
A more detailed comparison of $F_S$ with the perturbative expression \eqref{eq:Fsg4} at short distances is in Fig.~\ref{fig:Fs-short}.

As in the case of the Polyakov loop correlator, at very large distances the singlet correlator gets screened, 
first by the Debye mass, then by dynamically generated asymptotic screening masses~\cite{Bazavov:2018wmo,Steinbeisser:2018sde}.

\section{Conclusions}
In summary, the Polyakov loop has been computed up to order $g^5$, the (subtracted) $Q\bar{Q}$ free energy at short distances up to corrections of order $g^7(r\pi T)^4$, $g^8$, 
and the singlet free energy at short distances up to corrections of order $g^4(r\pi T)^5$, $g^6$.
Lattice calculations are consistent with weak-coupling expectations.
From the entropy of the Polyakov loop one can estimate a transition temperature to the quark-gluon plasma of $153^{+6.5}_{-5}$~MeV.

The comparison of thermal loop functions with weak-coupling calculations enables to address some fundamental questions,
like the set-in region, in terms of temperatures and distances, of screening.
An open theoretical question is if one of the free energies, notably the singlet one, may be related to the real-time potential. 
Finally, we have seen that effective field theories provide a proper definition of the $Q\bar{Q}$ color octet free energy at short distances, 
which can be tested and computed in lattice QCD.

\section*{Acknowledgements}
I would like to express my gratitude to Alexei Bazavov, Matthias Berwein, Nora Brambilla, Jacopo Ghiglieri, Peter Petreczky and Johannes Weber for collaborating 
at various stages and to various degrees on the work presented here. 
This work has been supported by the DFG and the NSFC through funds provided to the Sino-German CRC 110 ``Symmetries and the Emergence
of Structure in QCD'', and by the DFG cluster of excellence ``Origin and structure of the universe'' (www.universe-cluster.de). 

\bibliography{conf13}

\end{document}